\documentclass[aps,nofootinbib,preprint,tightenlines]{revtex4}

\usepackage{graphicx}
\usepackage{amsmath}
\usepackage{amssymb}
\usepackage{bm} 

\newcommand{\beq}{\begin{eqnarray}}
\newcommand{\eeq}{\end{eqnarray}}
\newcommand{\bqa}{\begin{eqnarray}}
\newcommand{\eqa}{\end{eqnarray}}

\newcommand{\sing}{^1{\rm S}_0}
\newcommand{\trip}{^3{\rm S}_1}

\newcommand{\e}{\mathrm{e}}
\newcommand{\dd}{\mathrm{d}}
\newcommand{\dde}{\frac{\mathrm{d}}{\mathrm{d}E}}
\newcommand{\halb}{\frac{1}{2}}
\newcommand{\lp}{{\ell^{\prime}}}
\newcommand{\lpp}{{\ell^{\prime\prime}}}
\renewcommand{\mp}{{m^{\prime}}}
\newcommand{\mpp}{{m^{\prime\prime}}}

\def\mqo2{{\!\!\!}}

\begin{document}


\preprint{HISKP-TH-08/17}
\title{Efimov physics in a finite volume}
\author{Simon Kreuzer}

\author{H.-W. Hammer}
\affiliation{Helmholtz-Institut f\"ur Strahlen- und Kernphysik (Theorie)
and Bethe Center for Theoretical Physics,
 Universit\"at Bonn, 53115 Bonn, Germany\\}

\date{\today}

\begin{abstract}
  Three bosons with large scattering length show universal properties
  that do not depend on the details of the interaction at short
  distances.  In the three-boson system, these properties include a
  geometric spectrum of shallow three-body states called \lq\lq Efimov
  states'' and log-periodic dependence of scattering observables on
  the scattering length.  We investigate the modification of the
  Efimov states in a finite cubic box and calculate the dependence of
  their energies on the box size using effective field theory.  
  We explicitly verify the renormalization of the effective field theory 
  in the finite volume.
\end{abstract}

\maketitle


Strongly interacting quantum systems can show universal properties that
are independent of the details of their interaction.  These properties 
establish connections between systems over a wide range of scales
and with different underlying interactions. One example
are the hydrodynamic properties of 
quantum liquids. The strongly interacting quark gluon plasma created at 
the Relativistic Heavy Ion Collider and ultracold Fermi gases both behave 
as nearly perfect liquids with almost no viscosity \cite{Cho08}.  
Their viscosity to entropy density ratio is close to a lower bound 
that was conjectured using string theory methods \cite{Kovtun:2004de}.

Another example are universal properties in non-relativistic few-body 
systems \cite{Braaten:2004rn}. 
Here the strong interaction regime is characterized by a
large scattering length $a$. If $a$ is large and positive, 
two particles of mass $m$ form a shallow dimer with energy
$E_2 \approx -{\hbar^2}/{(m a^2)}\,$,
independent of the mechanism responsible for the large scattering length.
Examples for such shallow dimer states are the deuteron 
in nuclear physics, the $^4$He dimer in atomic physics, and 
possibly the new charmonium state $X(3872)$ in particle 
physics \cite{Braaten:2004rn}.  These systems span more than 13 orders of
magnitude in energy ranging from MeV to neV.
In the three-body system, the universal properties include the
Efimov effect \cite{Efimov-70}.  If at least two of the three pairs of
particles have a large scattering length $|a|$ compared to the range
$r_0$ of their interaction, there is a sequence of three-body bound
states whose energies are spaced geometrically between
$-\hbar^2 / m r_0^2$ and $-\hbar^2 / m a^2$.  In the limit $1/a \to 0$, there
are infinitely many bound states with an accumulation point at
the three-body scattering threshold. These Efimov states or trimers have a
geometric spectrum \cite{Efimov-70}:
\begin{eqnarray}
E^{(n)}_3 = -(e^{-2\pi/s_0})^{n-n_*} \hbar^2 \kappa^2_* /m,
\label{kappa-star}
\end{eqnarray}
where $\kappa_*$ is the binding momentum of the Efimov trimer
labeled by $n_*$. This spectrum is a signature of a discrete scaling
symmetry with discrete scaling factor $e^{\pi/s_0}$.  In the case of
identical bosons, $s_0 \approx 1.00624$ and the discrete scaling
factor is $e^{\pi/s_0} \approx 22.7$.  This discrete
scale invariance is also relevant if $a$ is large but finite.
It becomes manifest in the log-periodic dependence of 
scattering observables on the scattering length $a$ \cite{Efimov79}.  The
consequences of discrete scale invariance and Efimov physics can be
calculated in an effective field theory for short-range interactions,
where the Efimov effect appears as a consequence of a renormalization
group limit cycle \cite{Bedaque:1998kg}.

Experimental evidence for an Efimov trimer in ultracold Cs atoms was
recently provided by their signature in three-body recombination rates
\cite{Kraemer-06}. This signature could be unravelled by varying the
scattering length $a$ over several orders of magnitude using a
Feshbach resonance.  More recently, evidence for Efimov trimers was
also obtained in atom-dimer scattering \cite{Knoop08} and possibly in
three-body recombination in a balanced mixture of atoms in three
different hyperfine states of $^6$Li \cite{Ottenstein08,Huckans08}.

The observation of Efimov physics in nuclear and particle physics
systems is complicated by the inability to vary the scattering length.
See, e.g., Ref.~\cite{Canham:2008jd} for
a recent discussion of Efimov physics in halo nuclei.  
Another opportunity to observe Efimov physics
is given by lattice QCD simulations of three-nucleon systems
\cite{Wilson:2004de}.  A number of studies of the quark-mass
dependence of the chiral nucleon-nucleon ($NN$) interaction found that
the inverse scattering lengths in the relevant $\trip$--$^3{\rm D}_1$
and $\sing$ channels may both vanish if one extrapolates away from the
physical values to slightly larger quark masses
\cite{Beane:2001bc,Beane:2002xf,Epelbaum:2002gb}.  This implies that
QCD is close to the critical trajectory for an infrared RG limit cycle
in the three-nucleon sector.  It was conjectured that QCD could be
tuned to lie precisely on the critical trajectory by tuning the up and
down quark masses separately \cite{Braaten:2003eu}.
As a consequence, the triton would display the Efimov effect.  More
refined studies of the signature of Efimov physics in this case
followed \cite{Epelbaum:2006jc,Hammer:2007kq}.  However, a proof of
this conjecture can only be given by an observation of this effect in a
lattice QCD simulation \cite{Wilson:2004de}.  The first full lattice
QCD calculation of nucleon-nucleon scattering was recently reported in
\cite{Beane:2006mx} but statistical noise presents a serious challenge
and no three-nucleon calculation has been carried out to date.  Since
lattice simulations are carried out in a cubic box, it is important to
understand the properties of Efimov states in the box.  Apart from
this application, the modification of Efimov physics in a finite
volume is an interesting question on its own and could be tested with
ultracold atoms in optical lattices.

The modifications to the Efimov spectrum can be calculated in
effective field theory (EFT) since the finite volume modifies the
infrared properties of the system. The properties of three-body
systems in a finite volume have been studied previously.  For
repulsive and weakly attractive interactions without bound states, Tan
has determined the volume dependence of the ground state energy of
three bosons up to~$\mathcal{O}((a/L)^7)$ \cite{Tan08}.  In
Refs.~\cite{Beane:2007qr,Detmold:2008gh}, this result was extended for
general systems of $N$ bosons. 
First results for three and more boson systems from lattice QCD have become 
available recently~\cite{Beane:2007es,Detmold:2008fn,Detmold:2008yn}.
For the unitary limit of infinite scattering length some studies
have been carried out as well.
The properties of three spin-1/2 fermions in a box were 
investigated in \cite{Pricoup07}.  In
this system, the Efimov effect does not occur due to the Pauli
principle.  The spectrum of three bosons in a harmonic trap was
calculated by Werner and Castin \cite{Werner06}.

In this letter, we investigate the finite volume corrections for a
three-boson system with large but finite scattering length.  In
particular, we derive an integral equation for the trimer energies in
a system of three identical bosons in a cubic box and solve this
equation for a few exemplary cases.  We follow the strategy proposed
in \cite{Beane:2003da} and extend the the EFT for the three-boson
system with short-range interactions of Ref.~\cite{Bedaque:1998kg} to
a cubic box.  We also verify explicitly that the renormalization of
the EFT is not affected by the box.
Our study is carried out to leading order in the large scattering 
length. This corresponds to the zero range limit with $r_0=0$. Our
results will be applicable to physical states whose size is large
compared to $r_0$. Using units with $\hbar=m=1$ from now on,
the effective Lagrangian can be written as \cite{Braaten:2004rn}
\begin{eqnarray}
\mathcal{L} &=&
\psi^\dagger \left(i \partial_t + \halb \nabla^2 \right) \psi
+ \frac{g_2}{4} d^\dagger d 
- \frac{g_2}{4}\left(d^\dagger\psi^2 + \mathrm{h.c.}\right)
- \frac{g_3}{36}d^\dagger d \psi^\dagger\psi +\ldots\,,
 \label{eq_lagrangian}
\end{eqnarray}
where the dots indicate higher order terms.
It involves the boson field~$\psi$ and an auxiliary dimer field $d$. 
The coupling 
constants $g_2$ and $g_3$ are matched to the scattering length 
and a three-body observable, respectively.

\begin{figure}[t]
	\centerline{
          \includegraphics*[width=8cm]{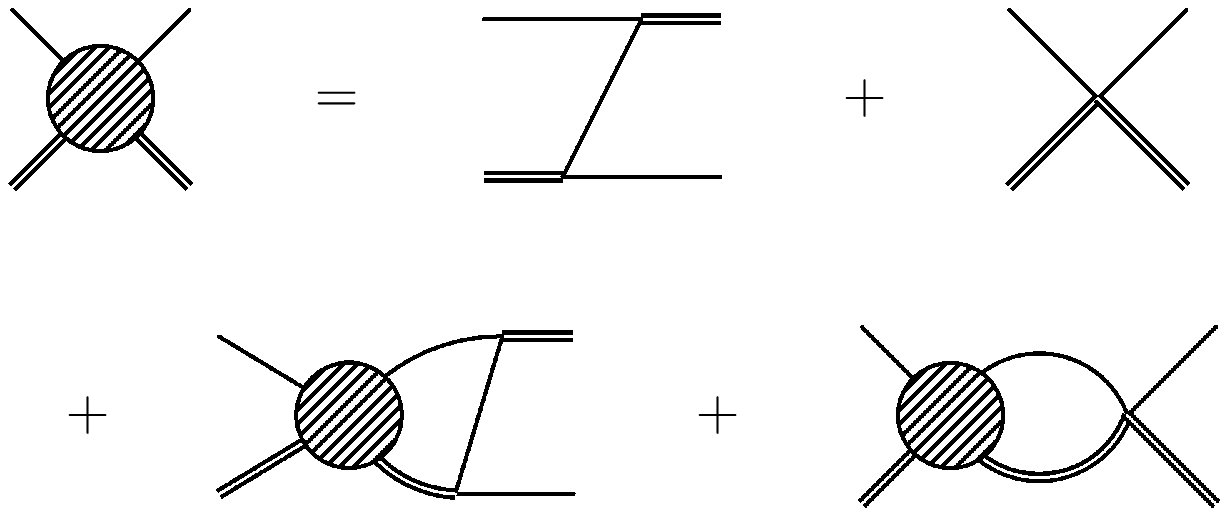}
        }
	\caption{Integral equation for the atom-dimer
          amplitude.  The single lines denote the boson propagator and
          the double lines denote the full dimer propagator.}
	\label{fig:inteq}
\end{figure}

The atom-dimer amplitude is determined by the inhomogeneous
integral equation depicted in Fig.~\ref{fig:inteq}, where the single lines
denote the boson propagator and the double lines denote the
full dimer propagator. 
In a cubic box, the momenta are quantized and the loop integrals in
the equation are replaced by discrete sums over quantized momenta
$\vec{k} = (2\pi/L)\,\vec{n},\,\vec{n}\in\mathbb{Z}^3$, where $L$ is
the side length of the box. The loop sums are regulated via a momentum
cutoff~$\Lambda$ just as in the infinite volume case.  All observables
are independent of $\Lambda$.  This is achieved by an appropriate
dependence of the coupling constant $g_3$ on $\Lambda$
\cite{Bedaque:1998kg}.  This dependence can be calculated
analytically, but one three-body input is required to fix the unknown
constant $\Lambda_\ast$ (see below). This renormalization procedure is
performed in the infinite volume limit.  The ultraviolet physics
associated with the renormalization of the EFT should not interfere
with the infrared physics brought about by putting the system into a
finite volume, as long as the corresponding scales are well
seperated. 
We will verify this expectation explicitly below.

The full dimer propagator $D$ denoted by the double lines
in Fig.~\ref{fig:inteq} is obtained by dressing the bare dimer
propagator which is simply a constant, $4i/g_2$,
with bosonic loops.
This leads to an infinite sum which can be evaluated analytically yielding
\begin{equation}
D(E,\vec{0}) = \frac{32\pi}{g_2^2} 
\Bigg[ \frac{1}{a} - \sqrt{-E} + \frac{1}{L}
\sum_{\vec{\jmath} \in \mathbb{Z}^3\atop\vec{\jmath}\neq 0} 
\frac{1}{|\vec{\jmath}|} \e^{-|\vec{\jmath}|\sqrt{-E}L} \Bigg]^{-1}
\label{dimprop}
\end{equation}
for a dimer at rest.
The term containing the box length $L$ vanishes in the limit
$L\to\infty$ and the expression reduces to the infinite
volume result.
Using the Feynman rules encoded in Eq.~(\ref{eq_lagrangian}) and the
full dimer propagator from above, we can translate the Feynman diagrams
in Fig.~\ref{fig:inteq} into an equation
for the atom-dimer amplitude. It involves an integration over the
loop energy and a sum over over the quantized loop momenta.
The integration over the loop energy is performed using the residue theorem
while the remaining
sum over the quantized momenta is rewritten into an integral by
virtue of Poisson's resummation formula:
$\sum_{\vec{n} \in \mathbb{Z}^3} \delta^3 (\vec{z}-\vec{n})
=\sum_{\vec{m} \in \mathbb{Z}^3} \exp(i2\pi \vec{m}\cdot\vec{z})\,$,
which is understood to be used under an integral.

The resulting expression is simplified further by exploiting the
behavior of the atom-dimer amplitude near a bound state. For
energies close to a bound state energy, the amplitude has a
simple pole, and the dependence on the ingoing and outgoing momenta
seperates.  Thus, we obtain a homogeneous integral equation for the
bound state amplitude~$\mathcal{F}$, namely
\begin{equation}
  \label{eq_hom_inteq}
  \mathcal{F}(\vec{p}) = \frac{1}{\pi^2} \int^\Lambda \dd^3 y
  \sum_{\vec{m} \in \mathbb{Z}^3} \e^{i L \vec{m}\cdot\vec{y}}
  \mathcal{Z}_E(\vec{p}, \vec{y})\tau_E(y) \mathcal{F}(\vec{y})\,,
\end{equation}
where
\begin{equation}
  \mathcal{Z}_E(\vec{p}, \vec{y}) =
  \left( p^2 + \vec{p}\cdot \vec{y} + y^2 - E \right)^{-1} 
          + \frac{H(\Lambda)}{\Lambda^2}\,,
\label{Beq_aux}
\end{equation}
\begin{equation}
  \tau_E(y) = \Bigg(-\frac{1}{a} + \sqrt{\frac{3}{4}y^2 - E} -
    \sum_{\vec{\jmath} \in \mathbb{Z}^3\atop\vec{\jmath}\neq \vec{0}}
    \frac{1}{L|\vec{\jmath}|}
    \e^{-L|\vec{\jmath}|\sqrt{\frac{3}{4}y^2 - E}} \Bigg)^{-1}\,.
\nonumber
\end{equation}
The  $\Lambda$-dependent three-body interaction in Eq.~(\ref{Beq_aux})
is given by
$H(\Lambda) \equiv -g_3\Lambda^2/(9g_2^2)=
\cos[s_0 \ln(\Lambda/\Lambda_*)+\arctan s_0]/
\cos[s_0 \ln(\Lambda/\Lambda_*)-\arctan s_0]\,$ where 
$s_0 \approx1.00624$ is a transcendental number \cite{Bedaque:1998kg}.

In the infinite volume case, only S-wave bound states are formed.
In a cubic box, however, the extraction of the S-wave
part of~$\mathcal{F}$
is not straightforward due to the breakdown of the spherical
symmetry to a cubic symmetry. The infinitely many irreducible 
representations of the spherical
symmetry are mapped onto the five irreducible representations of the cubic
symmetry. To investigate the mixing of the different partial waves,
the part of the amplitude with quantum number $\ell$ is projected out
\begin{equation}
\label{eq_inteq_blm}
\begin{split}
  F_{\ell}(p) c_{\ell 0} =&
      \frac{4}{\pi} \int_0^\Lambda \dd y\, y^2 \bigg[
          Z_E^{(\ell)}(p, y)\tau_E(y) \frac{c_{\ell 0}}{2\ell+1} F_\ell(y)
          +2\sqrt{\pi}
          \sum_{\vec{n} \in \mathbb{Z}^3\atop \vec{n} \neq \vec{0}}
          \sum_{\lp, \mp}^{(A_1)}\sum_{\lpp,\mpp}
          \begin{pmatrix}\lp & \lpp & \ell \\ 0 & 0 & 0\end{pmatrix}
          \begin{pmatrix}\lp & \lpp & \ell \\ \mp & \mpp & 0\end{pmatrix}\\
      &   \quad\times \sqrt{\frac{(2\lp+1)(2\lpp+1)}{2\ell+1}}\;i^\lpp
          j_{\lpp}(L|\vec{n}|y)
Y_{\lpp \mpp}(\Omega_{\vec{n}}) Z^{(\ell)}_E(p,y)
          \tau_E(y) c_{\lp\mp} F_{\lp}(y) \bigg]
\end{split}
\end{equation}
where we have used Wigner 3-$j$ symbols and
\begin{equation}
\frac{Z^{(\ell)}_E(p, y)}{2\ell+1} = 
\Bigg[\frac{1}{py}
    Q_\ell\left(\frac{p^2+y^2-E}{py}\right) + 
    \frac{H(\Lambda)}{\Lambda^2}\delta_{\ell 0} \Bigg]
\end{equation}
with $Q_\ell$ a Legendre function of the second kind. Moreover,
$j_{\lpp}$ is a spherical Bessel function of order~$\lpp$. 
The $\lpp$ sum runs over all partial waves while the $\lp$
sum runs over partial waves associated with the $A_1$ representation
of the cubic group, namely $\ell^\prime=0,4,6,\dots$.  The size of the
contribution of the respective partial wave to the cubic group
harmonic of the $A_1$ representation is given by the coefficients
$c_{\ell m}$~(See, e.g., Refs.~\cite{Bethe47,Altmann65}).
The first term in Eq.~\eqref{eq_inteq_blm} reproduces the equation for the
corresponding partial wave of the infinite volume amplitude, while the
second term yields corrections due to the quantization of the momenta
and admixtures of other partial waves.

For our study of the trimer states in the cubic box,
this equation is now specialized to the $\ell=0$ case:
\begin{eqnarray}
  \label{eq_inteq_b0}
  F_0(p) 
   &=& \frac{4}{\pi}\int_0^\Lambda \dd y\, y^2 Z^{(0)}_E(p, y) \tau_E(y) 
        \Bigg(1+\sum_{\vec{n}\in\mathbb{Z}^3\atop \vec{n}\neq\vec{0}}
       \frac{\sin(L|\vec{n}|y)}{L|\vec{n}|y}\Bigg)F_0(y) +\ldots\,,
\end{eqnarray}
where the dots indicate admixtures from higher partial waves.  Since
the leading term in the expansion of the Bessel functions in
Eq.~\eqref{eq_inteq_blm} is $1/(L|\vec{n}|y)$ these contributions are
suppressed by at least $a/L$. They will be small for volumes not too
small compared to the size of the bound state.  The lowest partial
wave that is mixed with the S-wave is the $\ell=4$ wave. Moreover,
contributions from higher partial waves will be suppressed
kinematically for shallow states with small binding momentum. This is
ensured by the spherical harmonic in the second term of
Eq.~\eqref{eq_inteq_blm}. Only for small lattices, i.~e. when $a/L$ is
large, this behavior is counteracted by terms stemming from the
spherical Bessel function $j_\ell(L|\vec{n}|y)$ and higher partial
waves may contribute significantly. We will therefore
only consider the terms explicitly given in Eq.~(\ref{eq_inteq_b0})
and leave the calculation of corrections due to higher partial waves
for future work.
For the numerical solution of Eq.~\eqref{eq_inteq_b0}, a set of basis
functions is chosen to turn the integral equation into a homogeneous
matrix equation. The basis functions chosen are shifted Legendre
polynomials defined on the intervall $[0, \Lambda]$ with a logarithmic
dependence in the argument. This behavior is tailored to the known
log-periodic behavior of the infinite volume bound state
amplitude~\cite{Bedaque:1998kg}. The set of basis functions is
orthonormal with respect to a suitable chosen scalar product. The
energy $E$ is now varied such that the resulting matrix has the
eigenvalue one. In order to reduce the numerical effort,
a Taylor expansion of the integral kernel 
\begin{equation}\label{eq_taylor}
 Z_E^{(0)}(p,y)\tau_E(y) = Z_{E_3^\infty}^{(0)}(p, y)\tau_{E_3^\infty}(y) +
 \dde\left.Z_E^{(0)}(p,y)\tau_E(y)\right|_{E=E_3^\infty}(E-E_3^\infty)+
 \mathcal{O}\left((E-E_3^\infty)^2\right)
\end{equation}
around the infinite volume energy~$E_3^\infty$ is performed.
In the following, we consider only the linear term in this expansion.
We expect this approximation to be valid for small energy shifts; its
validity will be discussed in detail below.
More details on the numerical method will be presented in a 
forthcoming publication.


\begin{figure}[t]
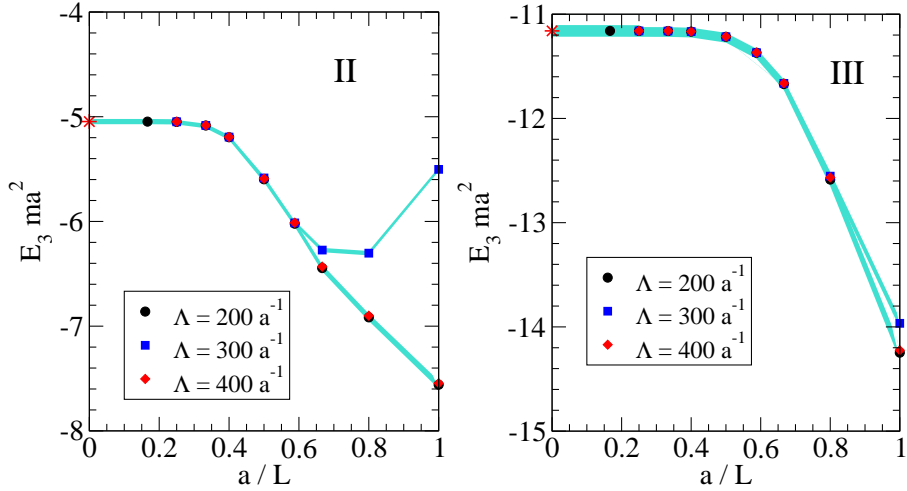

	\centerline{
          \includegraphics*[width=5.83cm]{new.eps}\ 
          \includegraphics*[width=6cm]{nunu.eps}
        }
	\caption{Variation of the trimer energy $E_3$ with
          the length $L$ of the cubic volume for the states II
          (left) and III (right). Plotted are three datasets for
          different values of the cutoff parameter $\Lambda$, together
          with the $1/(\Lambda a)$ bands. The point $a/L=0$ corresponds to 
          the infinite volume limit.}
	\label{fig:s23}
\end{figure}

For this first study of the trimer energies in a cubic box,
we take $a>0$ and
choose four states that have different energies in the infinite
volume including shallow as well as deep states:
\begin{itemize}
\item[Ia:] $E_3^\infty\, = -1.18907/a^2$, \quad $\Lambda_\ast a=5.66$,
\item[Ib:] $E_3^\infty\, = -27.4427/a^2$, \quad $\Lambda_\ast a=5.66$,
\item[II:] $E_3^\infty\, = -5.04626/a^2$, \quad $\Lambda_\ast a=1.66$,
\item[III:] $E_3^\infty\,= -11.1322/a^2$, \quad $\Lambda_\ast a=3.66$,
\end{itemize}
where $E_3^\infty$ is the trimer energy in the infinite volume.
Note that the states Ia and Ib appear in the same physical
system characterized by $\Lambda_\ast a=5.66$.

\begin{figure}[t]
	\centerline{
          \includegraphics*[width=9cm]{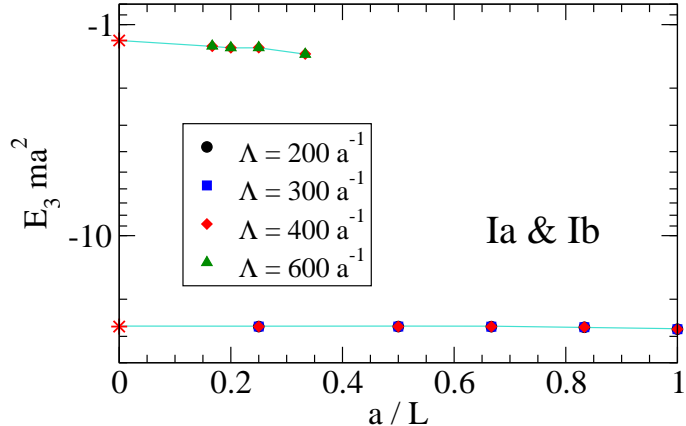}
        }
	\caption{Variation of the trimer energy $E_3$ with $L$ for the states Ia
          (upper curves) and Ib (lower curves). Plotted are two (three)
          datasets for different values of $\Lambda$, 
          together with the $1/(\Lambda a)$ bands.}
	\label{fig:s14}
\end{figure}

For each of these states, the energy in finite volume has been
calculated for various $L$.
Our results are shown in Figs.~\ref{fig:s23} and \ref{fig:s14}. 
In order to check the consistency of our results,
the calculation was carried out for several cutoff
momenta~$\Lambda$ and proper renormalization was verified explicitly.
For each cutoff the three-body interaction $H(\Lambda)$ was
adjusted to give the correct energy in infinite volume. 
The calculated values in Figs.~\ref{fig:s23} and \ref{fig:s14}
agree with each other very well
within the depicted error bands of the size $1/(\Lambda a)$
as long as the finite volume energy shift is not larger than 
15--20\% of the infinite volume energy.
These error bands give an estimate of the finite cutoff corrections
and renormalized quantities have to agree within these errors.
They do not correspond to the error from higher order corrections
in the EFT. 

When the finite volume shifts become larger than 15--20\%
of the infinite volume energy, the 
calculated results for different cutoffs in 
Fig.~\ref{fig:s23} start to disagree
and proper renormalization cannot be achieved anymore.
For the shallower state II this occurs for $a/L\approx 0.6$
while the deeper state III can be treated in the linear approximation
up to $a/L\approx 0.85$. 
This behavior of the energy shift for the shallow states
indicates the breakdown of the Taylor 
expansion of the integral kernel. As expected,
the linear approximation in Eq.~(\ref{eq_taylor}) is not applicable 
anymore if the finite volume shift becomes to large.
Calculations that do 
not rely on this expansion require a higher numerical effort. 
Such calculations are currently performed and the results will 
be presented in a forthcoming publication.

In Fig.~\ref{fig:s14}, we show our results for the two states Ia and Ib
in the same system characterized by $\Lambda_* a= 5.66$
for box sizes where the linear approximation in the kernel is
justified.
For the shallowest state~Ia reliable results can not be obtained for 
box sizes $L<3a$ while for the deepest state~Ib no 
appreciable finite volume shift can be detected for $L > a$.
Generally, the $L$ dependence of the energy shows a similar
behavior for all investigated states. For large box sizes, the finite
volume energy agrees with the infinite volume value. If the
box size is decreased, the binding increases.
The finite volume corrections are most important 
for the shallowest states which 
are largest in size and feel the finite volume first. 
For example, the ratio of the energies of the states Ia and Ib decreases
from 23.08 in the infinite volume limit to 19.5 for $L=3a$.
Note that this ratio differs from the discrete scaling factor 
$\exp(2\pi/s_0)\approx 515$ even in the infinite volume limit. 
This behavior is expected for shallow states close to the bound state 
threshold \cite{Braaten:2004rn}.
The ratio $\exp(2\pi/s_0)$ will be approached if deeper 
states are considered.

In this letter, we have studied the Efimov spectrum of
three bosons in a cubic box with periodic boundary conditions.
The knowledge of these modifications is important in order to
understand results from future 3-body lattice calculations. 
Using the framework of EFT, we derived a general equation for the 
trimer energies in the box. We solved this equation 
for box sizes as small as $L=a$ in a linear
approximation of the kernel for small finite volume shifts
and studied the modifications
of the spectrum as the system is squeezed into the box.
Moreover, we investigated the breakdown of the linear approximation.
We find that the expanded kernel seems to be applicable as long as the 
finite volume shift in the energy is not larger than 15--20\%
of the infinite volume energy. 
Numerically more expensive 
calculations using the unexpanded equation are under way.
As a next step, the role of the higher partial waves contributing to
the bound state amplitude has to be investigated in detail. This might 
allow to push the calculations to even smaller volumes and the regime
$L/a <1$ similar to the two-body system \cite{Beane:2003da}.
Finally, our method should be extended to the three-nucleon system
in order to be able to test the conjecture of Ref.~\cite{Braaten:2003eu}
in the future. This requires also the inclusion of higher
order corrections in the EFT and finite temperature effects
as lattice calculations will inevitably be performed at a small,
but non-zero, temperature.


This research was supported by the DFG through 
SFB/TR 16 \lq\lq Subnuclear structure of matter'' and the BMBF 
under contract No. 06BN411.



\end{document}